\documentclass[a4paper]{article}
\usepackage{adjustbox}
\usepackage{INTERSPEECH_v2}
\usepackage{url}
\title{Adversarial Network Bottleneck Features for Noise Robust Speaker Verification}
\name{Hong Yu$^{1,2}$, Zheng-Hua Tan$^2$, Zhanyu Ma$^1$, Jun Guo$^1$}
\address{
  $^1$Beijing University of Posts and Telecommunications, China\\
  $^2$Aalborg University, Denmark}
\email{hongyu@bupt.edu.cn, zt@es.aau.dk, mazhanyu@bupt.edu.cn, guojun@bupt.edu.cn }

\begin{document}

\maketitle
%
%
%
\begin{abstract}
In this paper, we propose a noise robust bottleneck feature representation which
is generated by an adversarial network (AN). The AN includes
two cascade connected networks, an encoding network (EN) and a discriminative network (DN).
Mel-frequency cepstral coefficients (MFCCs) of clean and noisy speech are
used as input to the EN and the output of
the EN is used as the noise robust feature.
The EN and DN are trained in turn, namely,
when training the DN, noise types are selected as the training
labels and when training the
EN, all labels are set as the same, i.e., the clean speech label, which
aims to make the AN features invariant to noise and thus achieve noise robustness.
We evaluate the performance of the proposed feature on a Gaussian Mixture Model-Universal Background Model based speaker
verification system,
and make comparison to MFCC features of speech enhanced by short-time
spectral amplitude minimum mean square error (STSA-MMSE) and deep neural
network-based speech enhancement (DNN-SE) methods.
Experimental results on the RSR2015 database show that the proposed AN bottleneck
feature (AN-BN) dramatically outperforms the STSA-MMSE and DNN-SE based MFCCs
for different noise types and signal-to-noise ratios.
Furthermore, the AN-BN feature is able to improve the speaker verification
performance under the clean condition.
\end{abstract}

\section{Introduction}
Recently, generative adversarial networks (GANs)~\cite{goodfellow2014generative} have attracted a tremendous amount of attention and they are successfully applied to many signal generation tasks, such as image generation~\cite{qi2017loss} and image to image translation~\cite{isola2016image}~\cite{radford2015unsupervised}~\cite{zhao2016energy}.
A GAN is composed of two networks: a generative network (GN) and a discriminative network (DN).
The GN is trained to generate 'fake' data from random inputs and make the generated 'fake' data similar to the 'real' data. The DN is trained to distinguish between the 'fake' and 'real' data.
By training these two networks in turn, the generated 'fake' data become more and more similar to the 'real' data. The GAN methodology is an instance of the broader machine learning concept called adversarial training, in which several networks learn together toward competing objectives, resulting in adversarial networks (ANs). An example application of ANs is dialogue generation \cite{li2017adversarial}.

So far in the area of audio and speech processing, ANs have received comparatively less attention than they have in image processing.  However,
some notable exceptions have been published recently. For example a phone/senone classifier is trained by adversarial learning methods in~\cite{serdyuk2016invariant}\cite{shinohara2016adversarial}, and an AN is used for music generation in~\cite{mogren2016c}.
%

In this work, we study ANs to address a well-known problem in speech processing, namely the significant degradation of performance of speech systems under noisy environments. In order to improve the robustness of these systems, in the literatures, a variety of speech enhancement methods are used to recover the clean speech signal from a noisy one, such as a priori Signal-to-noise ratio (SNR) estimation based Wiener filter~\cite{scalart1996speech}, short-time spectral amplitude minimum mean square error (STSA-MMSE)~\cite{erkelens2007minimum} and non-negative matrix factorization (NMF)~\cite{thomsen2016speaker}.
Many deep neural network (DNN) based methods have also been exploited.
In~\cite{plchot2016audio}\cite{xu2015regression}\cite{kolboek2016speech}, DNNs are used to enhance speech directly by obtaining
a denoised time-frequency representation. In~\cite{kolbaek2017speech}\cite{wang2014training}, an ideal time-frequency binary mask (IBM) or ideal time-frequency ratio mask (IRM) is estimated by DNNs firstly and is then used to recover clean speech.

In this paper, we propose a non-task-specific adversarial network for extracting bottleneck features (AN-BN).
Similar to GANs, the AN-BN extractor also includes two cascade connected networks, an encoding network (EN) and a discriminative network (DN).
Unlike GAN using random inputs, the AN uses clean and noisy acoustic features as training data and noise types as training labels.
%
The EN is trained to produce AN-BN features which are invariant to noise types and the DN is trained to distinguish the types of additive noises.
By training them in turn, noise robust AN-BN features are produced by the EN.
%

The proposed AN-BN features are applied to speaker verification (SV).
As we know, the performance of classical SV systems, such as Gaussian Mixture Model-Universal Background Model (GMM-UBM)~\cite{reynolds2000speaker} and i-Vector systems~\cite{dehak2011front},
greatly degrades when speech signals are corrupted by additive noises~\cite{yan2016improved}.
Many works have been done on developing noise robust SV systems during last decades~\cite{plchot2013developing}.
In the back end, pooling clean and noisy speech together to train SV systems is able to make the trained model better fit the noisy conditions~\cite{lei2013noise}\cite{lei2012towards}.
In the front end, a variety of speech enhancement methods, e.g., Wiener filter~\cite{scalart1996speech}, STSA-MMSE~\cite{erkelens2007minimum}and DNN speech enhancement~\cite{erkelens2007minimum}\cite{plchot2016audio}\cite{kolboek2016speech}\cite{kolbaek2017speech} are used. For the comparison purpose, the STSA-MMSE and DNN speech enhancement (DNN-SE) front ends are chosen as baseline front ends for a text-dependent SV system under different noise conditions.
%
%
%
%
%
%
%

The paper is organized as follows. In Section~\ref{sec_2}, we introduce the structure of the proposed AN-BN feature extractor and the training method.
In Section~\ref{sec_3}, we introduce two baseline frontends, STSA-MMSE and DNN-SE for the comparison purpose.
In Section~\ref{sec_4}, the speech corpora and noise data used for AN training and SV evaluation
are described.
In Section~\ref{sec_5}, the experimental design and results are presented, and finally the paper is concluded in Section~\ref{sec_6}.

\section{AN-BN feature extractor}
\label{sec_2}
\begin{figure}[t]
  \centering
  \includegraphics[width=\linewidth]{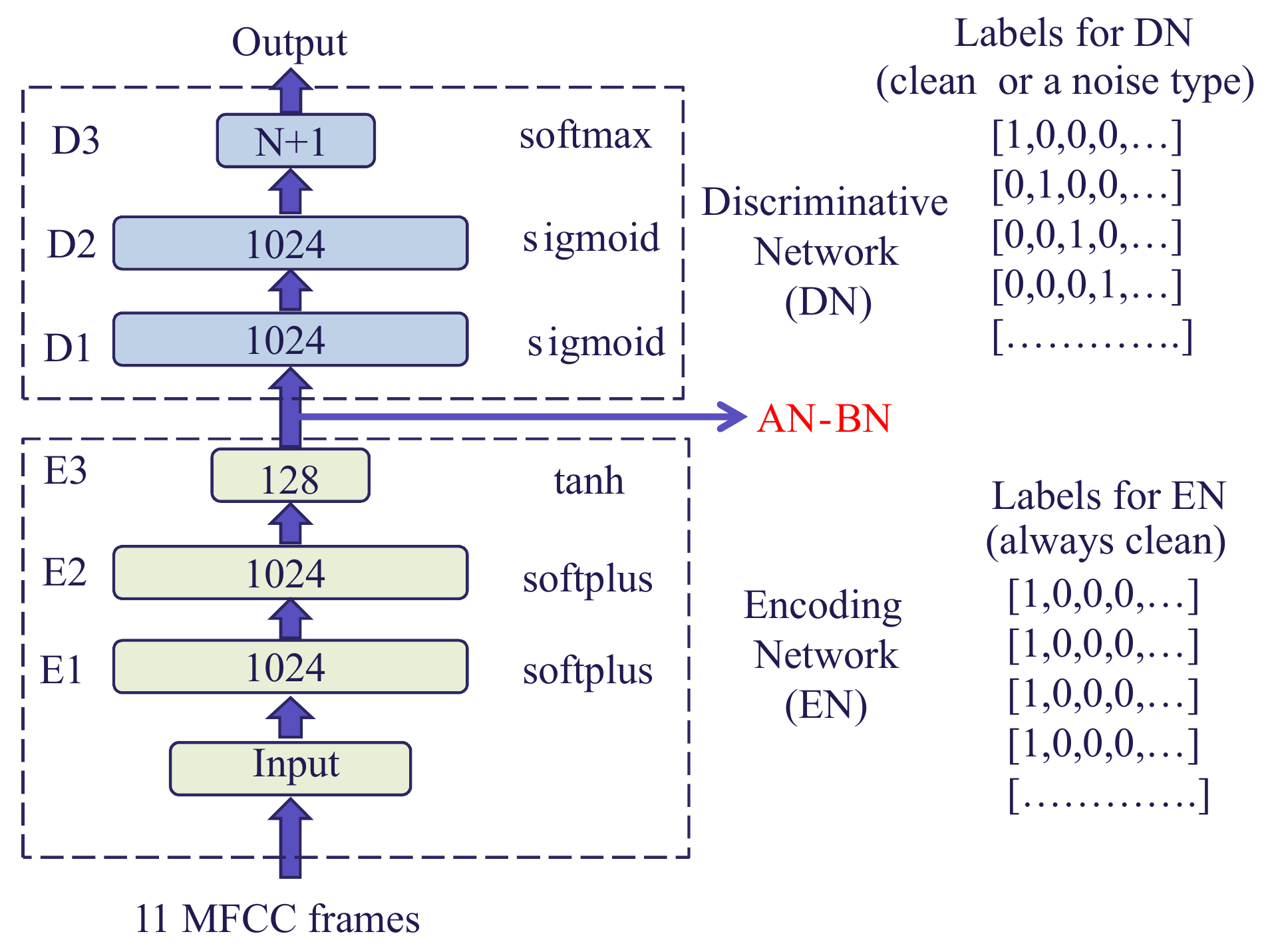}
  \caption{The structure of AN bottleneck feature extractor.}
  \label{fig:GAN}
\end{figure}

The proposed AN-BN feature extractor consists of two cascade connected networks, an EN and a DN, as shown in Fig.~\ref{fig:GAN}.
The EN includes three hidden layers, E1, E2 and E3, with 1024, 1024 and 128 nodes, respectively. Following the suggestion in~\cite{timmurphy}, the activation functions of E1 and E2 are both chosen as softplus ($\log(\exp(x)+1)$) and tanh is selected as the activation function of E3.

The input to the EN is batch normalized mel-frequency cepstral coefficients (MFCCs) of 11 frames including as context five past frames and five future frames, and the output of E3 is used as the AN-BN  feature. The DN includes two sigmoid hidden layers with 1024 nodes each and a softmax output layer. The dimension of the output layer is $N+1$, representing  $N$ noise types and clean.

%

When training the DN, noise types are used as training labels, and we update the parameters $\theta_D$ of the DN only, while keeping the values of parameters $\theta_E$ of the EN unchanged.
When training the EN, the label 'clean speech' is used for all inputs so as to output noise-invariant features, and we update $\theta_E$ only, while keeping the values of $\theta_D$ unchanged.

Clean and noisy training data are randomly grouped into small batches with 32 utterances each, and stochastic gradient descent (SGD) is used to train the EN and DN. The number of training epochs is selected as $30$.

The cross entropy function is selected as the cost function as shown in equations (\ref{eq1}) and (\ref{eq2}), where $x_i$ means the input feature, $m$ means the number of frames in each mini-batch, $L_{E_i}$ and $L_{D_i}$ stand for the training labels of the $i$-th frame, used for EN and DN training, respectively.
\begin{equation}
\footnotesize{
\min_{\theta_E}loss_{E}=-\frac{1}{m}\sum_{i=1}^{m}L_{E_i}\log[DN(EN(x_{i}))]^T,
  \label{eq1}}
\end{equation}
\begin{equation}
\footnotesize{
\min_{\theta_D}loss_{D}=-\frac{1}{m}\sum_{i=1}^{m}L_{D_i}\log[DN(EN(x_{i}))]^T.
  \label{eq2}
  }
\end{equation}
\section{Baseline systems}
\label{sec_3}
In this section, we introduce two baseline front-ends, STSA-MMSE and DNN-SE.
We also describe the GMM-UBM based SV baseline system which will be used to evaluate the performances of different front ends.
The GMM-UBM method is chosen as it performs well for short utterances~\cite{Delgado2016Asru}\cite{liu2015deep}, which is the case in this paper.
\subsection{STSA-MMSE}
STSA-MMSE is a noise independent speech enhancement method which does not need the apriori knowledge of noise type or noise level.
It is a statistical method which relies on the assumption that discrete Fourier transform (DFT) coefficients
of noise free speech follow a generalized gamma distribution~\cite{erkelens2007minimum}.
In the STSA-MMSE method, the priori SNR is estimated by the Decision-Directed approach~\cite{ephraim1984speech} and the noise power spectral density (PSD) is estimated by the noise PSD tracker reported in~\cite{hendriks2010mmse}. For each utterance, the noise tracker is initialized using a noise PSD estimate based on the first 1000 samples.
\subsection{DNN based speech enhancement}
The IRM estimation based DNN-SE method introduced in~\cite{kolbaek2017speech} is used as another baseline front-end. Following the suggestion in~\cite{kolbaek2017speech},
the time-frequency (T-F) representation used to construct the IRM is based on a gammatone filter bank with 64 filters linearly spaced on a Mel frequency scale and with a bandwidth
equal to one equivalent rectangular bandwidth (ERB)~\cite{wang2006computational}. The output of each filter bank channel is divided into 20 ms frames with 10 ms overlap.
IRM of noisy speech is used as the training label. On the $n$-th frame of channel $\omega$, IRM can  be computed as follows,
\begin{equation}
\mathbf{IRM}(n,\omega )=\left ( \frac{\left \|x(n,\omega)  \right \|^2}{\left \| x(n,\omega) \right \|^2+\left \| d(n,\omega) \right \|^2} \right )^{0.5},
  \label{eq3}
\end{equation}
where $\left \|x(n,\omega)  \right \|^2$ means the energy of clean speech of channel $\omega$ on the $n$-th frame  and $\left \|d(n,\omega)  \right \|^2$ stands for the energy of noisy speech of channel $\omega$ on the $n$-th frame. So the label dimension of each training feature frame is $64$.

The input to the DNN is a combination of features including 31 MFCCs, 15 amplitude modulation spectrogram (AMS), 13 relative spectral transform
perceptual linear prediction (RASTA-PLP) and 64 Gammatone filter bank energies (GFE).  Delta and double delta features are computed and a context of 2 past and 2 future
frames is utilized, so the dimension of training features is $(31+15+13+64)\times3\times5=1845$. All feature vectors are normalized to zero mean and
unit variance.

The DNN for IRM estimation includes three hidden layers of 1024 nodes. The activation functions for the hidden layer are rectified
linear units (ReLUs)~\cite{nair2010rectified} and a sigmoid function is for the output layer. The values of the parameters are updated using the SGD approach and the mean square error (MSE) is chosen as the cost function. The number of training epochs is selected as $30$.

The trained DNN is used to estimate IRM for test speech, and the estimated IRM is used to reconstruct the T-F representation of enhanced speech.
All T-F units in each frequency channel are then concatenated and all overlapping parts are summed. A time domain enhanced speech signal can be synthesized
by compensating for the different group delays in the different frequency channels and adding 64 frequency channels~\cite{wang2006computational}.

\subsection{Speaker verification systems}
In this paper we use the classical GMM-UBM SV method to evaluate the performance of three different front-ends.
The GMM-UBM based SV system is built and tested in three steps. First, a universal background GMM model (UBM) is trained
by an expectation-maximization algorithm using a large amount of general speech data.
Secondly, enrollment speaker GMMs are created using maximum-a-posteriori (MAP) adaptation of the UBM. Finally, the SV score of test speech is computed as the log-likelihood ratio between the claimed speaker's GMM and the UBM.
Usually, only clean or enhanced clean enrollment speech data are used for speaker model training. Motivated by the \emph{multi-condition} training method introduced in
\cite{lei2013noise}\cite{lei2012towards}\cite{yu2016effect}, we also investigate the performance of \emph{multi-condition} speaker models which are trained by enhanced clean and  noisy speech.
\section{Speech corpora and noise data}
\label{sec_4}
\subsection{Speech corpora}
4380 male speaker utterances from the TIMIT corpus~\cite{garofolo1993darpa} are used for UBM training.
The clean speech data used for training AN, DNN-SE and speaker models and for testing SV are all from RSR2015 corpus~\cite{larcher2014text} as detailed in Table~\ref{tab:RSR2015}.
\begin{table}[!htb]
\renewcommand{\arraystretch}{0.9}
\caption{{ Male-speaker speech used for training AN, DNN-SE and speaker models and for testing SV.}}
\label{tab:RSR2015}
\centerline{
\begin{tabular}{| c | c | c |c|}
\hline
   System          &Text ID. &Sess. ID &Sprk. ID \\
\hline
   AN-BN train            &2-30 &1,4,7 & 51-100 \\
   DNN-SE train &2-30 &1,4,7 & 51-100 \\
   Spkr. Model train   &1    &1,4,7 & 2-50\\
   SV test      &1    &2,3,5,6,8,9 & 2-50 \\
\hline
\end{tabular}
}
\end{table}

A text-dependent SV system is constructed for $49$ male speakers.
For training speaker models, text ID 1 and sessions 1, 4, and 7 from male speakers from $m002$ to $m050$ are selected, and for SV testing, sessions 2, 3, 5, 6, 8 and 9 are used.
There are in total $49\times6=294$ utterances used for testing and the trial protocol consists of $49\times294=14406$ trials.

The AN and DNN-SE model are trained using text IDs $2-30$ and sessions 1,4 and 7 from male speakers from $m051-m100$.

Speech used for  AN, DNN-SE model and speaker model training was recoded by Samsung Nexus smart phone. SV testing speech was recoded by Samsung Galaxy S and
a HTC Desire smart phone, which can make an unmatched microphone/recording setting.

\subsection{Noises and noisy speech}
In order to simulate the real-life speaker verification scenarios, we consider five different types of noises: Babble, Cantine, Market, Airplane and white Gaussian noise (White).
White was generated in MATLAB, Babble was made by adding 6 random speech samples from the Librispeech database~\cite{panayotov2015librispeech},
Cantine noises were recoded by the authors. Market and Airplane noises were collected by Fondazione Ugo Bordoni (FUB) and are available on request from the OCTAVE project~\cite{fuboctave}. All noise data are split into three non-overlapping parts for noisy speech generation, which are used in AN and DNN-SE model training, \emph{multi-condition} speaker model training and SV testing, respectively.

Noisy speech is created by taking out a random segment of noise which matches the length of the speech signal, scaling the amplitude of the noise segment to desired SNR levels, and adding it to the speech.
The scaling factor is calculated using the ITU speech voltmeter~\cite{recommendation2005191}.

\section{Experimental results and discussion}
\label{sec_5}

In order to evaluate the performance of the AN-BN feature for SV, six versions of AN-BN features are investigated: five noise specific AN-BN (NS-AN) features, one for each
noise type, and one noise general AN-BN (NG-AN) feature.
NS-ANs are trained by clean speech and one particular noisy speech and NG-AN is trained by a combination of clean and all five types of noisy speech.

MFCCs used for the AN training are generated using a 20ms frame length and 10ms frame shift.
Energy based voice activity detection (VAD) method is used to delete non-speech frames.
The dimension of MFCCs is 57 (without the 0-th coefficient, including static, delta and double delta features), so the input layer of the AN-BN extractor has $57\times11=627$ nodes.

Because the DN converges faster than the EN, in order to balance the training of EN and DN, the AN training uses noisy speech with high SNRs 10dB and 20dB, which can not be easily distinguished from clean speech.
Furthermore, in each mini-batch training, we update the EN three times and update the DN with a 50\% probability only.

The same as the AN-BN front end,  we also investigate five noise specific DNN-SE (NS-DNN) front ends and one noise general DNN-SE (NG-DNN) front end which are trained by one particular noisy speech and a combination of five types of noisy speech, respectively. Clean and corresponding noisy speech are used for computing labels for training. SNRs of noisy speech used for training DNN-SE models  are also 10dB and 20dB.

For evaluating the basic front end (no enhancement) and STSA-MMSE and DNN-SE front ends, MFCCs of 57 dimensions (the same as for AN training) are used for training and testing the SV systems. For the AN-BN front end, the SV system is trained and tested using AN-BN features with 128 dimensions. The mixture number of GMMs is chosen as 512.

SV systems built on different front ends are evaluated in different noise conditions with SNRs ranging from 0dB - 20dB. The system is also tested on the  enhanced clean speech
in order to investigate the effect of noise robust front ends on the noise free condition.

Firstly, we investigate the performance using \emph{clean} speaker models. For no enhancement front end, clean speech is used for training speaker models, and for other front ends,
enhanced clean speech is used, which means each \emph{clean} speaker model is trained by three utterances.  Equal error rates (EER) are used to evaluate the performances of different font ends and the results are shown in Table~\ref{tab:clean}.

It can be seen that AN-BN and DNN-SE front ends outperform the STSA-MMSE front end.
NS-AN and NG-AN front ends achieve the lowest EERs for the majority of the test conditions.
Comparing with the DNN-SE front ends, AN-BN front end can decrease average EERs by about 25\% on White and Babble noise and about 40\% on the other three noise types.
Especially, on SNRs from 0dB to 5dB which are not used for training DNN-SE and AN models, NS-AN and NG-AN perform much better than NS-DNN and NG-DNN, respectively.

Thereafter, we investigate the SV performances under the \emph{multi-condition} training framework.
For noise specific situations, enhanced clean speech and one type of enhanced noisy speech with SNR 10dB and 20dB are used for training speaker models, which means each speaker model is trained by nine utterances.
For noise general situation, enhanced clean speech and all five types of enhanced noisy speech with SNR 10dB and 20dB are used, each \emph{multi-condition} speaker model is trained by 33 utterances.
About no enhancement and STSA-MMSE front end, only noise specific situations are considered.

From the experimental results shown in Table~\ref{tab:mul}, it can be observed that \emph{multi-condition} trained speaker models can improve the performance of SV systems. AN-BN front ends still get the best results for most of the test conditions.

It is surprising to observe that the NG-AN front end outperforms NS-AN for several SNRs and noise types,
which means in \emph{multi-condition} SV systems, more speaker model training utterances can help the learned model to fit complex noisy environments and improve the robustness of SV systems.

It is also found that under high SNRs and clean conditions, the AN-BN front end performs much better than the DNN-SE front end.
A reasonable explanation to this is that, during the AN training, in the EN updating step,  \emph{clean} speech data from different sessions are all trained using the same 'clean' label.
It means the AN can extract not only the common information
between clean and noisy speech, but also that of of different \emph{clean} speech data.
The DNN-SE method, however, sets the training target as recovering the clean speech from the corresponding noisy speech, but it does not train on  \emph{clean} speech.
That is why EERs of the DNN-SE front end are very similar to the no enhancement front end on clean condition and the AN-BN front end is able to greatly improve the SV accuracy.
Generally, comparing with the DNN-SE front end, the AN-BN front end performs better for the SV task. The dimension of the AN-BN front end is 128 while that of the DNN-SE is 57, so the models for the AN-BN front end have a higher complexity. Future work includes reducing the dimension of the AN-BN front end to 57 for a fair comparison using principal component analysis or making the final output of the EN 57 dimensions.

\begin{table}
\renewcommand{\arraystretch}{0.9}
\caption{ EER ($\%$) of the SV system using different methods for different noise types and SNRs (dB) on clean speaker model.}
\label{tab:clean}
\begin{adjustbox}{max width=0.45\textwidth}
\begin{tabular}{| c | c | c | c| c|c| c| c|}
\hline
noise   &SNR  &No Enh. &MMSE &NS-DNN &NG-DNN & NS-AN & NG-AN\\
\hline
       & 00           & 45.90         & 30.95         & 39.46         & 40.14  & \textbf{25.69}         & 27.02\\
       & 05           & 43.20         & 21.17         & 20.75         & 21.77  & \textbf{17.01}         & 17.81\\
       & 10           & 34.61         & 13.95  & \textbf{9.86}         & 10.88         & 10.24         & 11.35\\
White         & 15           & 26.28         & 10.20         & 7.82         & 8.16  & \textbf{6.48}         & 7.51\\
       & 20           & 16.91         & 8.50         & 6.12         & 6.80  & \textbf{4.42}         & 5.29\\
       & clean           & 6.99         & 5.80         & 6.02         & 5.67         & 3.84  & \textbf{3.41}\\
\cline{2-8}
       & mean           & 28.98         & 15.10         & 15.01         & 15.57  & \textbf{11.28}         & 12.07\\
\hline
       & 00           & 19.05         & 29.04         & 17.01  & \textbf{16.67}         & 19.03         & 17.87\\
       & 05           & 14.63         & 20.40         & 10.54         & 10.39         & 10.20  & \textbf{9.86}\\
       & 10           & 11.69         & 12.59         & 7.82         & 7.50  & \textbf{5.44}  & \textbf{5.44}\\
Babble         & 15           & 11.04         & 7.82         & 6.46         & 6.34         & 3.21  & \textbf{3.06}\\
       & 20           & 9.18         & 6.29         & 6.12         & 5.78  & \textbf{3.06}         & 3.40\\
       & clean           & 6.99         & 5.80         & 5.78         & 5.67  & \textbf{3.00}         & 3.41\\
\cline{2-8}
       & mean           & 12.10         & 13.66         & 8.96         & 8.73         & 7.32  & \textbf{7.17}\\
\hline
       & 00           & 20.72         & 19.09         & 18.71         & 19.94  & \textbf{9.18}         & 9.81\\
       & 05           & 19.20         & 12.37         & 8.58         & 9.18  & \textbf{5.10}         & 5.86\\
       & 10           & 14.74         & 8.16         & 6.12         & 6.12  & \textbf{3.60}         & 4.44\\
Cantine         & 15           & 11.81         & 6.80         & 5.49         & 5.78         & 3.40  & \textbf{3.06}\\
       & 20           & 8.50         & 6.12         & 5.31         & 5.44  & \textbf{3.25}         & 3.40\\
       & clean           & 6.99         & 5.80         & 5.10         & 5.67         & 4.08  & \textbf{3.41}\\
\cline{2-8}
       & mean           & 13.66         & 9.72         & 8.22         & 8.69  & \textbf{4.77}         & 5.00\\
\hline
       & 00           & 29.40         & 25.51         & 21.43         & 21.77  & \textbf{14.29}         & 14.43\\
       & 05           & 20.07         & 17.35         & 9.86         & 10.59  & \textbf{6.80}         & 7.82\\
       & 10           & 15.00         & 11.90         & 6.88         & 7.48  & \textbf{4.08}         & 4.42\\
Market         & 15           & 11.96         & 8.28         & 6.46         & 6.22  & \textbf{3.06}         & 3.64\\
       & 20           & 8.93         & 7.35         & 5.78         & 5.76  & \textbf{3.13}         & 3.40\\
       & clean           & 6.99         & 5.80         & 5.92         & 5.67         & 3.74  & \textbf{3.41}\\
\cline{2-8}
       & mean           & 15.39         & 12.70         & 9.39         & 9.58  & \textbf{5.85}         & 6.19\\
\hline
       & 00           & 21.09         & 17.69         & 16.99         & 15.99  & \textbf{9.86}  & \textbf{9.86}\\
       & 05           & 15.99         & 12.58         & 10.55         & 8.99  & \textbf{6.12}         & 6.46\\
       & 10           & 13.61         & 8.17         & 7.48         & 6.12  & \textbf{4.35}         & 5.10\\
Airplane         & 15           & 11.66         & 6.53         & 6.99         & 6.12  & \textbf{3.74}         & 4.08\\
       & 20           & 9.18         & 6.27         & 6.15         & 5.58  & \textbf{3.06}         & 3.63\\
       & clean           & 6.99         & 5.80         & 6.12         & 5.67  & \textbf{3.40}         & 3.41\\
\cline{2-8}
       & mean           & 13.08         & 9.51         & 9.05         & 8.08  & \textbf{5.09}         & 5.42\\
\hline
\end{tabular}
\end{adjustbox}
\end{table}
\begin{table}
\renewcommand{\arraystretch}{0.9}
\caption{ EER ($\%$) of the SV system using different methods for different noise types and SNRs (dB) on multi-condition speaker model.}
\label{tab:mul}
\begin{adjustbox}{max width=0.45\textwidth}
\begin{tabular}{| c | c | c | c| c|c| c| c|}
\hline
noise   &SNR  &No Enh. &MMSE &NS-DNN &NG-DNN & NS-AN & NG-AN\\
\hline
       & 00           & 35.88         & 30.95         & 27.21         & 26.19  & \textbf{22.04}         & 26.87\\
       & 05           & 24.40         & 20.07  & \textbf{9.52}         & 11.22         & 13.95         & 17.69\\
       & 10           & 18.37         & 7.48  & \textbf{6.12}         & 7.14         & 8.42         & 11.19\\
White         & 15           & 15.81         & 6.46  & \textbf{5.02}         & 5.10         & 5.80         & 7.50\\
       & 20           & 14.97         & 6.46         & 4.65  & \textbf{4.08}         & 4.42         & 5.10\\
       & clean           & 5.85         & 4.76         & 5.78         & 4.00         & 2.04  & \textbf{1.33}\\
\cline{2-8}
       & mean           & 19.21         & 12.70         & 9.72         & 9.62  & \textbf{9.44}         & 11.61\\
\hline
       & 00           & 21.77         & 33.50         & 16.26  & \textbf{16.00}         & 17.35         & 18.71\\
       & 05           & 15.37         & 23.13         & 9.52  & \textbf{9.18}         & 10.48         & 9.86\\
       & 10           & 11.93         & 16.23         & 6.99         & 5.44         & 5.27  & \textbf{5.13}\\
Babble         & 15           & 9.52         & 12.63         & 6.08         & 4.76  & \textbf{3.06}         & 3.39\\
       & 20           & 8.16         & 8.84         & 5.78         & 4.08         & 2.72  & \textbf{2.61}\\
       & clean           & 6.12         & 7.12         & 5.17         & 4.00         & 2.38  & \textbf{1.33}\\
\cline{2-8}
       & mean           & 12.15         & 16.91         & 8.30         & 7.19         & 6.88  & \textbf{6.84}\\
\hline
       & 00           & 24.11         & 19.05         & 12.93         & 11.61  & \textbf{8.77}         & 10.20\\
       & 05           & 17.22         & 12.59         & 5.91         & 5.78         & 5.78  & \textbf{5.44}\\
       & 10           & 12.93         & 8.21         & 4.42         & 5.10         & 4.10  & \textbf{3.75}\\
Cantine         & 15           & 10.88         & 6.91         & 4.25         & 4.57         & 3.74  & \textbf{3.40}\\
       & 20           & 9.18         & 6.12         & 4.27         & 4.08         & 3.59  & \textbf{2.38}\\
       & clean           & 7.48         & 6.32         & 3.78         & 4.00         & 2.50  & \textbf{1.33}\\
\cline{2-8}
       & mean           & 13.63         & 9.87         & 5.93         & 5.86         & 4.75  & \textbf{4.42}\\
\hline
       & 00           & 36.05         & 29.25         & 19.33         & 18.37  & \textbf{15.31}  & \textbf{15.31}\\
       & 05           & 26.06         & 21.07  & \textbf{8.16}  & \textbf{8.16}         & 8.50         & 8.50\\
       & 10           & 18.37         & 13.95         & 6.24         & 5.78  & \textbf{5.29}  & \textbf{5.29}\\
Market         & 15           & 13.32         & 10.98         & 5.41         & 4.44         & 3.88  & \textbf{3.64}\\
       & 20           & 9.18         & 7.82         & 4.53         & 4.42         & 3.06  & \textbf{2.72}\\
       & clean           & 5.44         & 6.76         & 4.29         & 4.00         & 2.29  & \textbf{1.33}\\
\cline{2-8}
       & mean           & 18.07         & 14.97         & 7.99         & 7.53         & 6.39  & \textbf{6.13}\\
\hline
       & 00           & 32.28         & 25.51         & 14.78         & 11.38         & 11.56  & \textbf{10.54}\\
       & 05           & 26.87         & 15.48         & 8.26  & \textbf{6.12}         & 7.69         & 6.46\\
       & 10           & 21.10         & 8.16         & 5.44  & \textbf{4.78}         & 6.11         & 4.95\\
Airplane         & 15           & 16.38         & 6.12         & 5.53         & 4.72         & 4.42  & \textbf{4.08}\\
       & 20           & 9.86         & 5.44         & 4.76         & 4.23         & 3.40  & \textbf{3.00}\\
       & clean           & 5.83         & 5.44         & 4.76         & 4.00         & 2.32  & \textbf{1.33}\\
\cline{2-8}
       & mean           & 18.72         & 11.03         & 7.26         & 5.87         & 5.92  & \textbf{5.06}\\
\hline
\end{tabular}
\end{adjustbox}
\end{table}
\section{Conclusions}
\label{sec_6}
In this paper, we proposed a new adversarial networks (AN) based noise robust feature extractor, which consists of two cascade connected networks, one encoding network (EN) and one discriminative network (DN).
The EN and DN are trained in turn and the outputs of the EN are used as robust features for speaker verification (SV).
When training the DN, the values of EN parameters are kept unchanged and noise types are used as training labels. When the EN is trained, the values of DN parameters are kept unchanged and all input speech data are assigned the same label, namely the clean speech label.
Being trained using clean and noisy speech, the AN bottleneck (AN-BN) features can not only gain the common information between noisy and clean speech, the common information among clean speech recoded in
different sessions can also be extracted. This trait makes the AN-BN features particularly suitable for the noise roust SV task.
Experimental results on the RSR2015 data base show that the AN-BN front end outperforms short-time spectral amplitude minimum mean square error (STSA-MMSE) and deep neural network base speech enhancement (DNN-SE) front ends for the majority of the tested conditions, especially on high signal to noise ratios (SNR) and clean conditions. In the future, we will conduct more extensive comparison with existing methods and evaluate the performance
of the AN-BN features on other speech applications under noisy conditions, e.g., speech recognition and spoofing detection.

\section{Acknowledgements}
This work is partly supported by the NNSF of China under No. 61402047,61273217,61401259, Beijing
Nova Program under No.Z171100001117049. Beijing National Science Foundation
No. 4162044, Scientific Research Foundation for Returned Scholars, Ministry
of Education of China, Chinese 111 program of Advanced Intelligence,
Network Service Grant B08004. OCTAVE Project (No. 647850), funded by
the Research European Agency (REA) of the European Commission, in its
framework programme Horizon 2020.
The authors would like to thank Morten Kolb{\oe}k  for his assistance and software used for the speaker verification and DNN speech enhancement baseline systems.

\newpage

\bibliographystyle{IEEEtran}

\bibliography{mybib}


\end{document}